\documentclass[aps,prl,twocolumn,reprint,superscriptaddress]{revtex4-1}

\usepackage{graphicx}
\usepackage{dcolumn}
\usepackage{bm}
\usepackage{amssymb}
\usepackage{amsmath}
\usepackage{subfigure}
\usepackage{epstopdf}
\usepackage{url}
\usepackage[breaklinks]{hyperref}
\usepackage{breakurl}

\newif\ifhyper
\hypertrue
\ifhyper
\hypersetup{
   citecolor = {green},
   colorlinks = {true},
   urlcolor = {blue}
}
\fi

\usepackage{color}
\usepackage{soul}

\def\BState{\State\hskip-\ALG@thistlm}

\begin{document}

\title{Retrieving Quantum Information with Active Learning}

\author{Yongcheng Ding}
\thanks{jonzen.ding@gmail.com}
\affiliation{International Center of Quantum Artificial Intelligence for Science and Technology (QuArtist) \\ and Department of Physics, Shanghai University, 200444 Shanghai, China}
\affiliation{Department of Physical Chemistry, University of the Basque Country UPV/EHU, Apartado 644, 48080 Bilbao, Spain}
\author{Jos\'e D. Mart\'in-Guerrero}
\thanks{jose.d.martin@uv.es}
\affiliation{IDAL, Electronic Engineering Department, University of Valencia,
Avinguda Universitat s/n, 46100 Burjassot, Valencia, Spain}
\author{Mikel Sanz}
\affiliation{Department of Physical Chemistry, University of the Basque Country UPV/EHU, Apartado 644, 48080 Bilbao, Spain}
\author{Rafael Magdalena-Benedicto}
\affiliation{IDAL, Electronic Engineering Department, University of Valencia,
Avinguda Universitat s/n, 46100 Burjassot, Valencia, Spain}
\author{Xi Chen}
\thanks{xchen@shu.edu.cn}
\affiliation{International Center of Quantum Artificial Intelligence for Science and Technology (QuArtist) \\ and Department of Physics, Shanghai University, 200444 Shanghai, China}
\affiliation{Department of Physical Chemistry, University of the Basque Country UPV/EHU, Apartado 644, 48080 Bilbao, Spain}
\author{Enrique Solano}
\thanks{enr.solano@gmail.com}
\affiliation{International Center of Quantum Artificial Intelligence for Science and Technology (QuArtist) \\ and Department of Physics, Shanghai University, 200444 Shanghai, China}
\affiliation{Department of Physical Chemistry, University of the Basque Country UPV/EHU, Apartado 644, 48080 Bilbao, Spain}
\affiliation{IKERBASQUE, Basque Foundation for Science, Maria Diaz de Haro 3, 48013 Bilbao, Spain}
\affiliation{IQM, Munich, Germany}

\date{\today}

\begin{abstract}
Active learning is a machine learning method aiming at optimal design for model training. At variance with supervised learning, which labels all samples, active learning provides an improved model by labeling samples with maximal uncertainty according to the estimation model. Here, we propose the use of active learning for efficient quantum information retrieval, which is a crucial task in the design of quantum experiments. Meanwhile, when dealing with large data output, we employ active learning for the sake of classification with minimal cost in fidelity loss. Indeed, labeling only 5\% samples, we achieve almost 90\% rate estimation. The introduction of active learning methods in the data analysis of quantum experiments will enhance applications of quantum technologies.

\end{abstract}

\maketitle
\textit{Introduction.---} In the past decades, machine learning has evolved from (un)supervised learning algorithms~\cite{unsupervised,supervised,semi}, aiming at simple classification tasks, to deep learning algorithms~\cite{hintondeep,lecundeep}, such as playing Go~\cite{alphago} and StarCraft \uppercase\expandafter{\romannumeral2\relax}~\cite{starcraft}. Supervised learning can lead to well-trained classification or prediction models by tuning them with labeled data. However, most data are unlabeled in real world, thus the cost of labeling can be critical in chemistry or biology experiments, destructive testing in industry, among others~\cite{ar:sverchkov2017,ar:tuia2011}. At the same time, machine learning protocols have shown their capabilities to attain quantum tasks and study properties of quantum systems~\cite{ar:Carleo2017,ar:Saito2017,ar:prl250501,ar:prl250502,ar:prl250503,ar:prb214306}. These protocols have already been applied in the field of quantum metrology, which is related to quantum information retrieval, making use of reinforcement learning (RL)~\cite{bk:sutton1998} to control certain aspects of the measurement process~\cite{inp:sanders2016, ar:tiersch2015}. We can also find quantum versions of RL in the scientific literature~\cite{ar:dong2008} for measurement control~\cite{ar:albarran2018, ar:cardenas2018}. The crucial problem of quantum information retrieval is the design of an optimal plan that minimizes the cost of measurements, while extracting the relevant  information for further tasks without well-defined rewards. Active learning (AL) is based on the hypothesis that a model trained on a small set of labeled samples can perform as well as one trained on a data set where all samples are labeled~\cite{inp:Baldridge2004, tr:settles2010}. Therefore, this framework fits well with the necessary requirement to address the aforementioned crucial information problem. In a nutshell, AL takes into account the cost of labeling, i.e. fidelity loss caused by measurement. It analyzes the most informative patterns (quantum states) in order to propose the minimal number of labels (measurements) which guarantee the maximal knowledge gain. There are recent works suggesting applications of AL to quantum information~\cite{ar:melnikov2018}, employing a definition of AL which is different to ours, assisting experimental design like other machine learning algorithms~\cite{stateoverlap,network}. An opposite approach is proposed in Ref.~\cite{qal}, which aims at accelerating classical AL by quantum computation. 

In this Letter, we propose a framework for making decisions about the optimal experimental design for binary classification with AL algorithms. For achieving this task, estimation models are updated in each iteration after labeling the qubit with the maximum uncertainty by means of weak measurements. These allow for the extraction of partial information while perturbing qubits slightly, implying cost reduction in the sense of fidelity loss. In our numerical simulations, we have observed that, by labeling only 5\% samples, we attain almost 90\% rate estimation for the task. We consider that the introduction of AL algorithms into experimental design could lead to improved applications in quantum technologies.

\textit{Active learning.---} Let be a set of labeled samples $X=\{\mathbf{x_i},y_i\}_{i=1}^l$, where the inputs $\mathbf{x_i}\in \mathcal{X}$, being $\mathcal{X}$ defined in $\mathbb{C}^d$, and for the sake of simplicity we consider a classification problem where the output is given by the corresponding class, $y_i\in\{1, \ldots, C\}$ for a $C$-class problem. To complete the definition of the AL framework, we also need a set of unlabeled samples $U=\{\mathbf{x_i}\}_{i=l+1}^{l+u} \in \mathcal{X}$, being $u \gg \ l$, i.e., the pool of candidates to be labeled is in principle much larger than those samples already labeled. AL usually works following an iterative procedure so that samples are labeled sequentially to improve the model performance. This is done by adding the most informative sample in each iteration up to a point where adding more labels do not benefit the model and, hence, the model can work on a semisupervised fashion using only the labeled samples. The obvious question is which are the most informative samples that should be selected. The usual approach is to consider that samples with maximal information are those for which the model displays maximal uncertainty about the outcome. Therefore, labeling these sample provides a considerable added value to the learning process. There are different approaches to evaluate the uncertainty in order to sort the samples in $U$ and make a decision about which candidate should be part of the training set. The two most widely used strategies are uncertainty sampling (USAMP) and query-by-committee (QBC)~\cite{inp:settles.nips08ws}. USAMP uses a single model for selecting samples with maximal uncertainty according to the estimator, and updates the model~\cite{inp:Lewis1994}. QBC employs several models to select for labeling the samples with the lowest consensus measured by voting entropy~\cite{inp:seung1992}.

For the simplest USAMP, assuming a probabilistic binary classification model, the strategy queries the sample whose conditional probability of being positive is nearest 0.5. When three or more classes are present, the criterion is to take the sample whose prediction is the least confidence
\begin{eqnarray}
x_{\text{LC}}=\underset{x}{\text{argmax}}(1-P_{\theta}(\hat{y} \arrowvert x)), \\
\hat{y}=\underset{y}{\text{argmax}}(P_{\theta}(y \arrowvert x)), \nonumber
\end{eqnarray}
with $\hat{y}$ the most probable class according to model $\theta$. Beyond this criterion, there are other approaches like margin sampling~\cite{inp:scheffer2001}, entropy-based USAMP~\cite{ar:shannon1948}, which differs in probability densities (see the Supplementary Material~\cite{sm}). We only introduce the least confidence sampling since these three approaches are the same when dealing with binary classification. Meanwhile, voting entropy for QBC, which considers the most informative sample, is defined by
\begin{eqnarray}
x_{\text{VE}}=\underset{x}{\text{argmax}}\left(- \sum_i \frac{V(y_i)}{C} \log \frac{V(y_i)}{C} \right),
\end{eqnarray}
where $y_i$ refers to all possible labelings, $V (y_i)$ is the number of votes received by the label from the members of the committee, and $C$ is the committee size. Alternative QBC approaches are also described in Supplementary Material~\cite{sm}.

\textit{Weak measurement.---} An extension of von Neumann measurement was proposed to extract information from a quantum system without destroying its quantumness, which is called weak measurement~\cite{ar:aharonov1964,ar:aharonov1987,ar:aharonov1988,ar:aharonov2005}. In our framework, the protocol of weak measurement consists of two steps: coupling the quantum system to an ancilla qubit for obtaining a new system, then followed by a projective measurement on the ancilla qubit. Let us suppose that the ancilla qubit's Gaussian wave function reads as
\begin{eqnarray}
|\Phi\rangle=\int\frac{1}{(2\pi\sigma^2)^{\frac{1}{4}}}\exp\left(-\frac{q^2}{4\sigma^2}\right)|q\rangle dq,
\end{eqnarray}
where $\sigma$ is the standard deviation of the qubit's position, $\hat{q}$ is the position operator of the qubit that $\hat{q}|q\rangle=q|q\rangle$. Accordingly, there exists the conjugate momentum operator $\hat{p}$ that satisfies the commutation relation $[\hat{q},\hat{p}]=i\hbar$. The ancilla qubit is coupled to the system following an interaction Hamiltonian
\begin{equation}
H_I(t)=g(t)\hat{p}\otimes\hat{A},
\end{equation}
where $g(t)$ is a time-dependent coupling strength, $\hat{A}$ is the operator of the quantity we aim to measure with eigenvectors $|a_j\rangle$ satisfying $\hat{A}|a_j\rangle=a_j|a_j\rangle$. We require the momentum of the ancilla qubit to be sufficiently small, which leads a small uncertainty in momentum and a large one in its position $q$. The time-dependent coupling strength now satisfies
\begin{equation}
\int_0^{t_0}g(t)dt=1,
\end{equation}
Therefore, the strength of the measurement is no longer governed by a coupling constant. Now the initial quantum state of the quantum system is $|\Phi\rangle\otimes|\Psi\rangle$, which evolves under the interaction Hamiltonian by $\exp[-i\int_0^tH_I(t')dt']$ ($\hbar=1$). One can see that within $t_0$, the interaction Hamiltonian takes $\hat{q}$ to $\hat{q}+a_j$ on each of the entangled wave functions of the detector and eigenvector of quantity to be measured $|\Psi\rangle\otimes|a_j\rangle$,
\begin{eqnarray}
\hat{q}(t_0)-\hat{q}(0)=\int_0^{t_0}dt\frac{\partial\hat{q}}{\partial t}=i\int_0^{t_0}[H_I,\hat{q}]dt=a_j.
\end{eqnarray}
The evolution of the wave function can be written as
\begin{eqnarray}
&\exp(-i\hat{p}\otimes\hat{A})|\Phi(q)\rangle\otimes|\Psi\rangle=\nonumber\\
&\cos\frac{\alpha}{2}|\Phi(q-a_1)\rangle\otimes|a_1\rangle+\sin\frac{\alpha}{2}|\Phi(q-a_2)\rangle\otimes|a_2\rangle.
\end{eqnarray}
In this way, we can obtain $\hat{A}$ of the qubit by measuring the ancilla's position $q$ with an arbitrary uncertainty, since weak measurement protocol requires $\sigma\gg\max_j(a_j)$. The probability distribution of the ancillary position gives
\begin{eqnarray}
&P(q)={(2\pi\sigma^2)^{-\frac{1}{2}}}\left[\cos^2\frac{\alpha}{2}\exp\left(-\frac{(q-a_1)^2}{2\sigma^2}\right)\right.\nonumber\\
&\left.+\sin^2\frac{\alpha}{2}\exp\left(-\frac{(q-a_2)^2}{2\sigma^2}\right)\right].
\end{eqnarray}
If we perform a weak measurement on the $Z$ direction of the qubit, $\hat{A}=\hat{\sigma_z}$, which leads to $|a_1\rangle=|0\rangle$, $|a_2\rangle=|1\rangle$, and $a_1,a_2=\pm1$, the probability $P(q)$ can be approximated by
\begin{equation}
P(q)\approx\frac{1}{(2\pi\sigma^2)^{\frac{1}{2}}}\exp\left[-\frac{(q-\cos\alpha)^2}{2\sigma^2}\right].
\end{equation}
A normalized wave function of the system after a quantum measurement on the ancilla is
\begin{eqnarray}
|\Psi_f\rangle\propto\frac{1}{(2\pi\sigma^2)^{\frac{1}{4}}}\left\{\cos\frac{\alpha}{2}\exp\left[-\frac{(q_0-1)^2}{4\sigma^2}\right]|0\rangle\right.\nonumber\\ 
\left.+\sin\frac{\alpha}{2}\exp\left[-\frac{(q_0+1)^2}{4\sigma^2}\right]|1\rangle\right\},
\end{eqnarray}
where $q_0$ is the measurement feedback of the ancilla position. The wave function $|\Psi_f\rangle$ is close to the initial wave function $|\Psi\rangle$ if $\sigma$ is large enough; i.e., the qubit is not destroyed but slightly perturbed. Although the weak measurement protects the qubit from collapsing, less information is extracted from the system than that from a direct measurement due to the uncertainty, which also increases the error of labeling. This trade-off is inevitable when we only have one qubit of $|\Psi\rangle$, but the error rate of labeling can be reduced if we introduce extra resources. For instance, if there are $n$ qubits prepared in the same state $|\Psi\rangle$ as an ensemble, the uncertainty of $\langle\hat{A}\rangle$ can be reduced by $1/\sqrt{n}$.

\begin{figure}
\includegraphics[width=8.6cm]{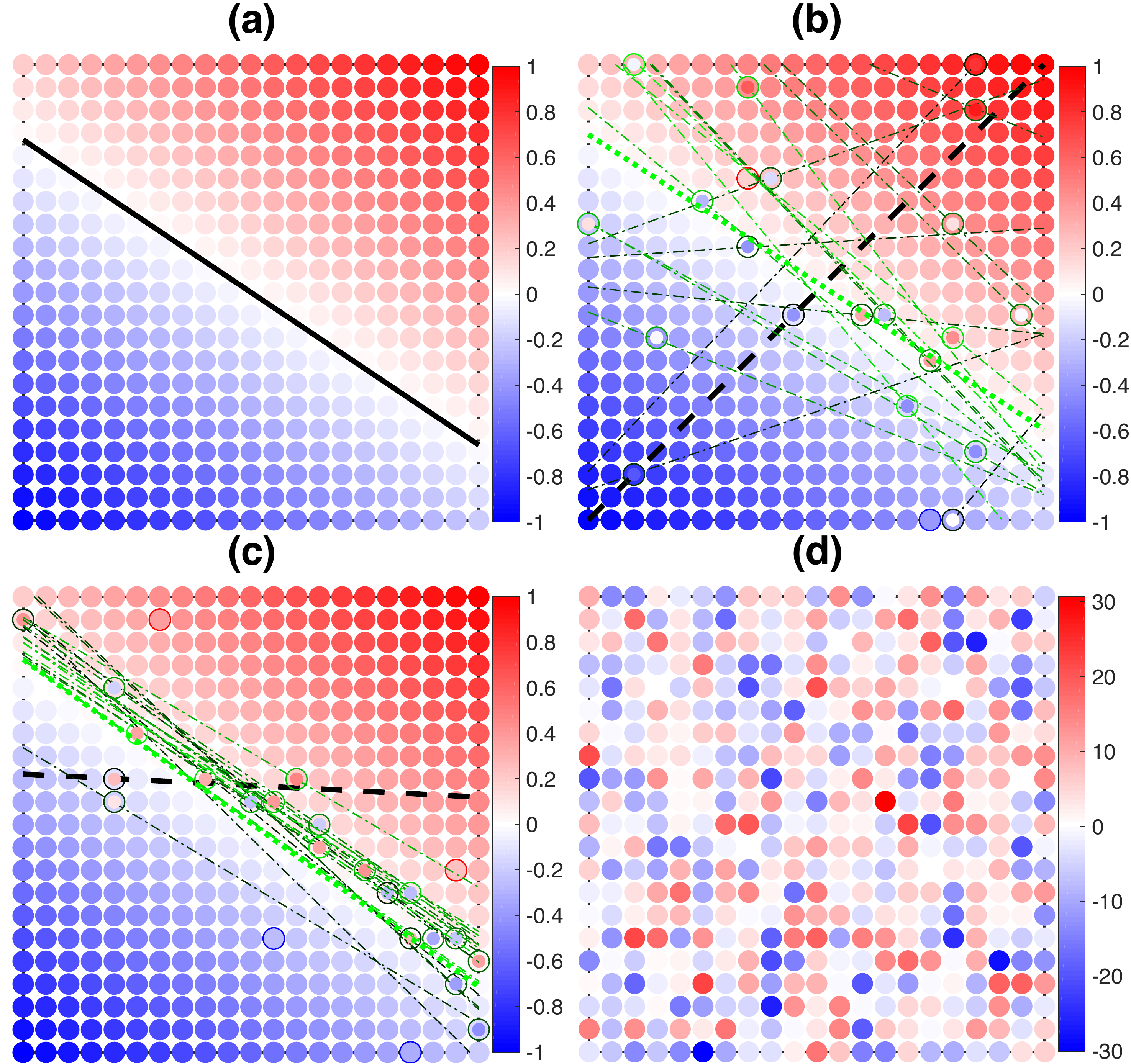}
\caption{\label{fig:alicebob}(a) The quantum state in a lattice of $21\times21=441$ qubits, prepared by Alice for binary classification. The value of $\langle\hat{\sigma_z}\rangle=\cos\alpha\in[-1,1]$ is plotted in the color map from blue ($-1$) to red (1). (b) USAMP protocol. Thick black dashed line represents the initial SVM that divides the lattice into two parts, using standardized support vectors of two oracles provided by Alice (circled by red and blue). Thin dash-dotted lines with colors from black to green illustrate the update of the model, where candidates which are selected according to USAMP strategy are circled in the same color. Qubits with the minimal fidelities among their $n$ copies are identified by Gaussian weak measurements with $\sigma=10$ and $n=500$. We have covered their initial states by smaller circles in different color, depending on the outcome. Thick green dotted line represents the SVM after labeling 20 samples via weak measurements. (c) QBC protocol. We present the evolution of the SVM as one model in our committee, where other parameters are unchanged. (d) Weak values of all qubits after performing weak measurements only once on the quantum state. These weak values contain little information which is hardly useful for classification, which requires $n$ copies of the quantum state for obtaining statistically meaningful information.}
\end{figure}

\textit{Numerical simulations.---} Here, we exemplify AL to a binary classification problem for quantum information retrieval. In Fig.~\ref{fig:alicebob}(a), Alice prepares a quantum state in a lattice of $21 \times 21 =$ 441 qubits, which can be mapped to a spin system with transformation $|0\rangle\rightarrow|\uparrow\rangle$ and $|1\rangle\rightarrow|\downarrow\rangle$. Information for classification can be encoded in $\hat{\sigma_z}$, e.g. $\langle\hat{\sigma_z}\rangle>0$ for class 0 and $\langle\hat{\sigma_z}\rangle<0$ for class~1. $n$ copies of the quantum system with qubits correctly labeled by Alice, which we call oracles, are sent to Bob for classification. Suppose Bob knows that the quantum system can be modeled linearly, the first trial is training a support vector machine (SVM) by USAMP with two oracles of different labels [see Fig.~\ref{fig:alicebob}(b)]. We label a candidate $\textbf{x}$, selected among other unlabeled samples based on its uncertainty, i.e., its effective distance to the current hyperplane. 
A more complex AL protocol based on QBC is shown in Fig.~\ref{fig:alicebob}(c), comprising a committee made up of four models: SVM, coarse Gaussian SVM, fine decision tree, and linear discriminant. Hence, Bob inquires about more oracles since the committee needs more information for minimal modeling. After a first round of evaluating the disagreement by voting entropy, candidate is selected according to the same rule as in USAMP among other samples with maximal committee disagreement. Different from classical labeling, we have a high error rate when we label a sample by weak measurement, since the protocol requires an inaccurate ancilla with large $\sigma$. In Fig.~\ref{fig:alicebob}(d), we plot the weak value of each qubit after performing weak measurements on the quantum system. One should average weak values of $n$ copies for obtaining meaningful information to reduce uncertainty, allowing us to correctly label each sample.

\begin{figure}
\includegraphics[width=8.6cm]{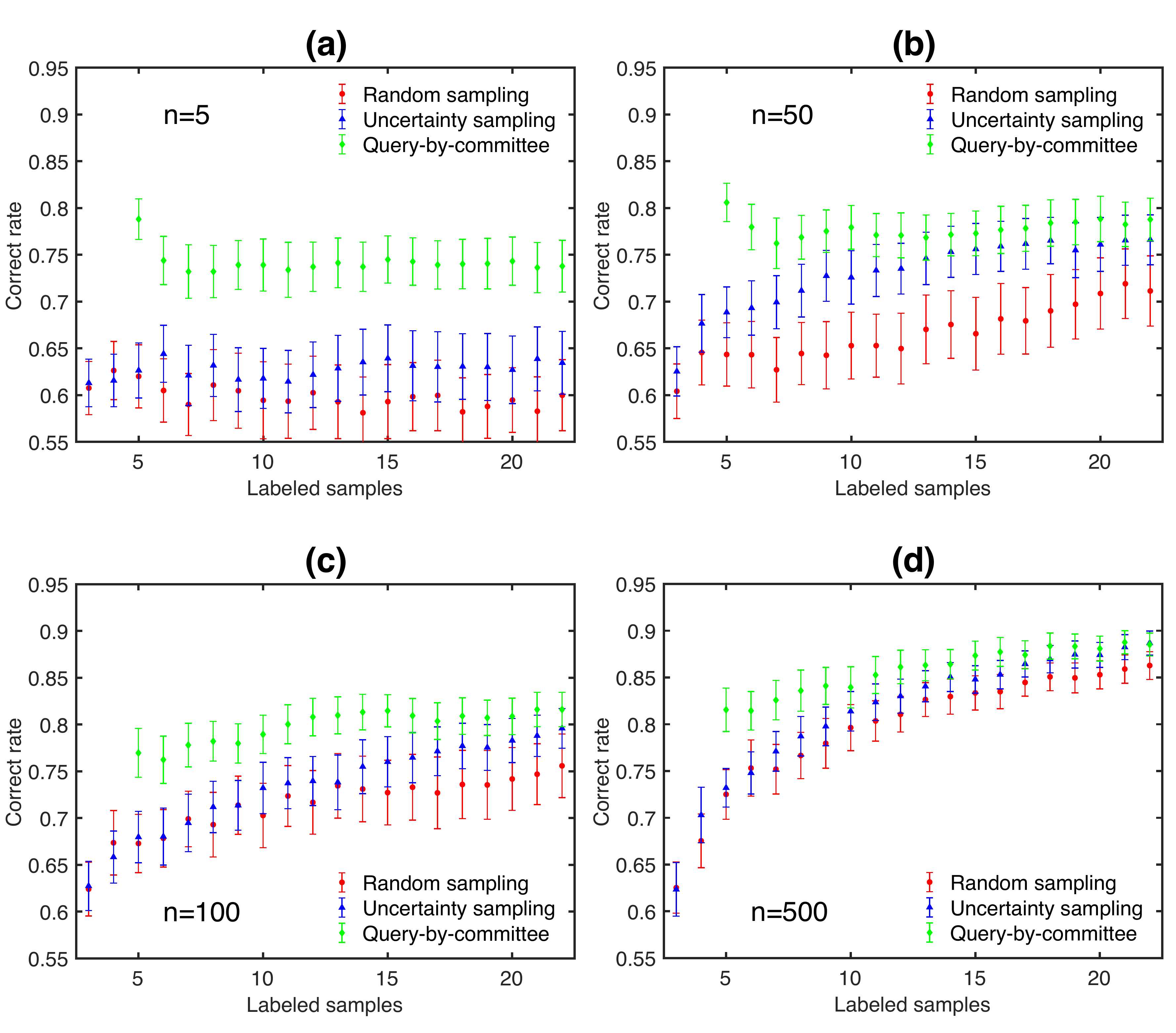}
\caption{\label{fig:three}Mean correct rates of classification model (SVM) with random sampling (red dots), USAMP (blue triangles) and QBC (green diamonds) as different sampling strategies. Error bars denote confidence intervals of 0.95. Each qubit is sampled over an ensemble of (a) 5, (b) 50, (c) 100, and (d) 500 qubits. Other parameters agree with those in Fig.~\ref{fig:alicebob}.}
\end{figure}

\begin{figure}
\includegraphics[width=8.6cm]{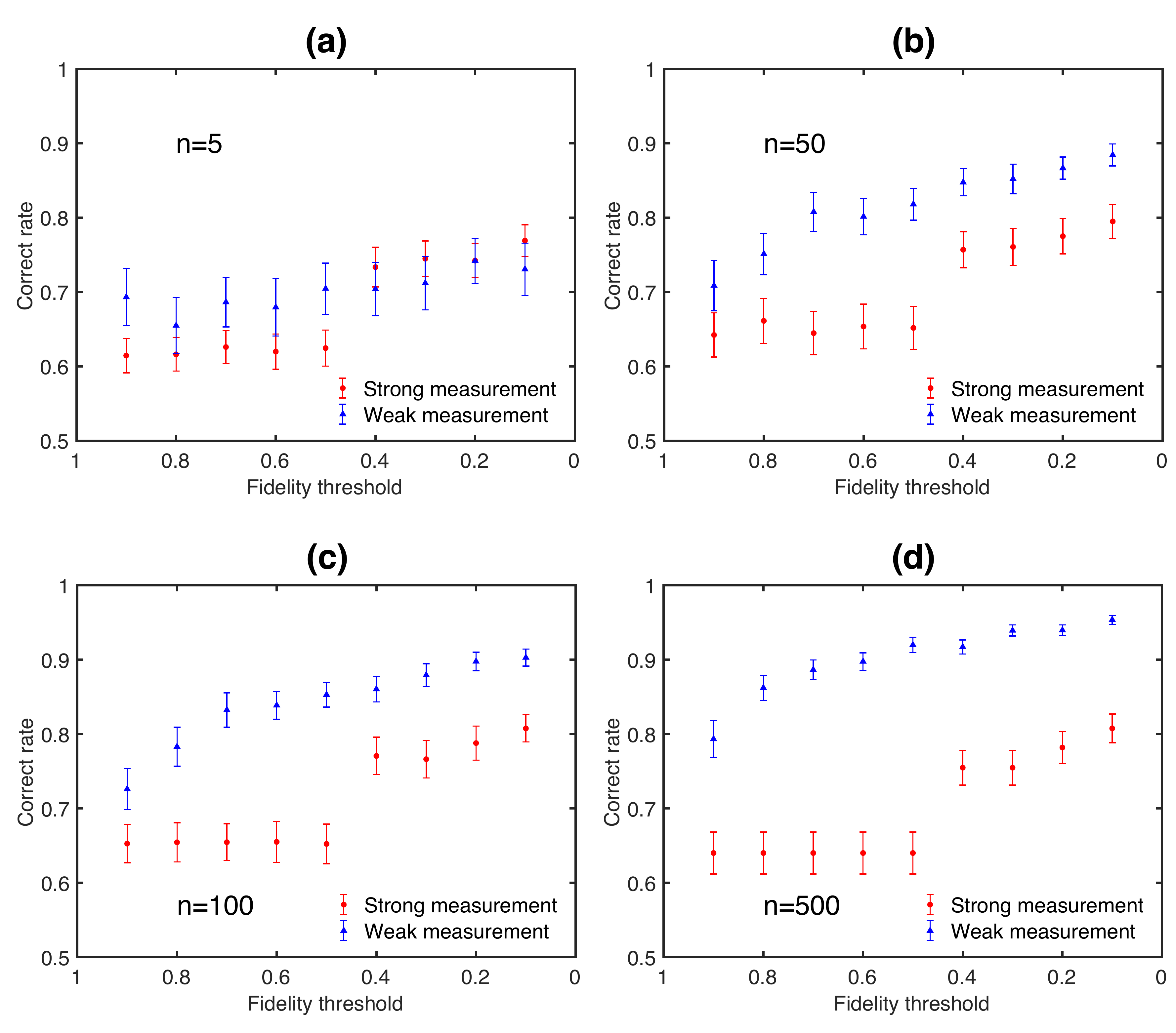}
\caption{\label{fig:two}Mean correct rates of classification model (SVM) with USAMP. Each qubit is labeled by strong measurement (red dots) and weak measurement (blue triangles). Parameters remain the same as in previous figures.}
\end{figure}

Now we present a more quantitative study by defining the cost of labeling in quantum measurement by fidelity loss. Once we fix the number of samples to be labeled or the fidelity threshold, different sampling strategies and measurement methods can be fairly compared. Here we evaluate the performance of every classification model by their rate estimation since the classes are balanced. One may use other figure-of-merits, e.g., AUC or ROC when they are unbalanced. In Fig.~\ref{fig:three}, we compare USAMP and QBC, which are the two most widely accepted strategies, against random sampling. The experiment starts with three labeled samples for USAMP and five for QBC. Result indicates that with an adequate choice of committee, QBC can be more efficient than USAMP since its correct rate is higher under different number of labeled samples. 
We also notice the anomaly that, under small $n$, QBC outperforms other methods with fewer labeled samples. This is because the training set consists of four correctly labeled oracles from Alice and samples labeled by Bob, which can be incorrectly labeled with a high probability when $n$ is small. This phenomenon becomes trifling when $n$ is sufficient large, as depicted in Fig.~\ref{fig:three}(d)]; i.e., almost every sample is faultlessly labeled. In Fig.~\ref{fig:two}, we compare strong and weak measurement in AL with USAMP under different fidelity thresholds. We measure each sample for updating our model until the fidelity of the system reaches the threshold. Weak measurements allow us to label more samples than strong measurement. For the calculation of fidelity loss, we multiply the state fidelity by the minimal fidelity of each labeled qubit for its $n$ copies after measurements. Meanwhile, a smaller $n$ might also enlarge the training set because a large fidelity loss is less likely to happen. This situation refers to a trade-off between information increment due to more samples and higher accuracy per sample in AL.

\textit{Conclusion and outlook.---} We have introduced AL protocols for retrieving quantum information with optimal experimental design. Moreover, we have exemplified with a complete binary classification task by extracting information from qubits through weak measurements. Furthermore, we have compared selection strategies using USAMP, QBC and random sampling, as well as labeling techniques employing weak and strong measurements. For the former, the results of our numerical simulations have shown that, with only 5\% of labeled samples, we have achieved almost 90\% rate estimation. We have observed that weak measurement strategy outperformed
strong measurement. Our framework includes the concept of trade-off and dynamical prediction, where its efficiency could be related to a generative model~\cite{DGM}. A straightforward extension of this work will be solving multiclass classification problem on qudits, where other approaches for USAMP such as margin sampling or entropy based sampling are no longer equivalent to least confidence. Another potential candidate platform for applications is quantum memristors~\cite{ar:pfeiffer,sc,photonics}, since they are based on the weak measurement protocol that allows feedback for controlling its coupling to the environment. An AL-enhanced quantum memristor could be a more efficient building block for quantum simulations of non-Markovian systems or neuromorphic quantum computation.

We acknowledge support from National Natural Science Foundation of China (NSFC) (Grant No. 11474193), STCSM (Grant No. 18010500400 and No. 18ZR1415500), Program for Eastern Scholar, Ram\'on y Cajal program of the Spanish MCIU (Grant No. RYC-2017-22482), QMiCS (Grant No. 820505) and OpenSuperQ (Grant No. 820363) of the EU Flagship on Quantum Technologies, Spanish Government PGC2018-095113-B-I00 (MCIU/AEI/FEDER, UE), Basque Government IT986-16, as well as the and EU FET Open Grant Quromorphic. This work is supported by the U.S. Department of Energy, Office of Science, Office of Advanced Scientific Computing Research (ASCR) quantum algorithm teams program, under field work Proposal No. ERKJ333.

\onecolumngrid
\appendix
\section{Supplementary Materials}

\section{Background}
Uncertainty sampling (USamp) as the most intuitive and widely used strategy~\cite{inp:Lewis1994} can be understood as follows. Let us assume a probabilistic classification model for the sake of simplicity. If the classification were binary, the strategy would query the sample whose posterior probability of being positive is nearest 0.5. When three or more classes are present, the criterion is to take the sample whose prediction is the least confidence
\begin{eqnarray}
x_{\text{LC}}=\underset{x}{\text{argmax}}(1-P_{\theta}(\hat{y} \arrowvert x)), \\
\hat{y}=\underset{y}{\text{argmax}}(P_{\theta}(y \arrowvert x)), \nonumber
\end{eqnarray}
thus being $\hat{y}$ the most probable class according to model $\theta$. As the criterion for the least confidence strategy only considers the most probable label, it may be losing information present in the other labels. A more general approach is given by the so-called margin sampling~\cite{inp:scheffer2001}
\begin{eqnarray}
x_{\text{M}}=\underset{x}{\text{argmin}}(P_{\theta}(\hat{y_1} \arrowvert x) - P_{\theta}(\hat{y_2} \arrowvert x)),
\end{eqnarray}
where $\hat{y_1}$ and $\hat{y_2}$ are, respectively, the first and second most probable class labels according to model $\theta$. The idea behind this approach is that samples separated by large margins are easy to classify, and hence, the most difficult sample to classify is given by small margins. Therefore, the latter is the most informative sample because knowing the true label would be the most valuable added value to discriminate between classes. The obvious limitation of the approach is that it only considers information about the two most probable classes. In order to take information about all classes, entropy can be used~\cite{ar:shannon1948}
\begin{eqnarray}
x_{\text{H}}=\underset{x}{\text{argmax}}\left(- \sum_i P_{\theta}(y_i \arrowvert x) \log P_{\theta}(y_i \arrowvert x)\right),
\end{eqnarray}
where $y_i$ refers to all possible labelings. Entropy measures the amount of information needed to represent a given information, and hence, it is usually considered as an assessment of the uncertainty or impurity in ML applications. Obviously, when dealing with binary classification, entropy-based USamp is equivalent to the margin and least confidence approaches
 Query-by-committee (QBC) is also based on uncertainties, but from a different perspective~\cite{inp:seung1992}. In this case, the main idea is to set a committee of models (usually called expert committee) and to focus on those samples for which there is a high discrepancy among the different models. Therefore, QBC will label that sample that creates most disagreement among the models of the committee; obviously a key issue here is to come up with a metric to evaluate the discrepancy. By way of example, if in a classification problem in which the different models find a sample equally probable according to a majority-voting strategy to belong to the different classes, then that will be the sample to be labeled. This approach is especially interesting when the different models actually represent different areas of the data space; the more different and the more disjointly complementary they are, the better. This is in general true for the performance of the committee but it is remarkably relevant to implement a sound QBC-based AL. Although this may suggest that large committees should be preferred, small committees formed by just two or three models have also reported good results~\cite{inp:settles.nips08ws}. There are many methods to measure the disagreement among models; the two most common methods are vote entropy and Kullback-Leibler (KL) divergence.
The former considers that the most informative sample is
\begin{eqnarray}
x_{\text{VE}}=\underset{x}{\text{argmax}}\left(- \sum_i \frac{V(y_i)}{C} \log \frac{V(y_i)}{C} \right),
\end{eqnarray}
where $y_i$ refers to all possible labelings, $V (y_i)$ is the number of votes received by the label for the members of the committee, and $C$ is the committee size. The sample to be labeled according to KL divergence is given by
\begin{eqnarray}
x_{\text{KL}}=\underset{x}{\text{argmax}}\left( \frac{1}{C} \sum_{c=1}^C D(P_{\theta^{(c)}}\|P_\mathcal{C})  \right),\\
D(P_{\theta^{(c)}}\|P_\mathcal{C})= \sum_i P_{\theta^{(c)}} (y_i|x) \log \frac{P_{\theta^{(c)}} (y_i|x)}{P_\mathcal{C} (y_i|x)},
\end{eqnarray}
where $\theta^{(c)}$ represents a model of the committee and $\mathcal{C}$ the whole committee, hence $P_\mathcal{C} (y_i|x)=\frac{1}{C} \sum_{c=1}^C P_{\theta^{(c)}} (y_i|x)$ gives the probability of $y_i$ being the correct label. KL divergence looks for the largest average difference between the label distributions of a singular committee member and the consensus. Meanwhile, many other AL approaches have been proposed, always with the goal of labeling the most informative samples. Two of the most interesting proposals are based on analyzing the expected model change, the expected error reduction and the variance reduction. Methods based on density analysis also deserve being mentioned. Expected model change is based on labeling the sample that produces the greatest change in the model. There are many different possibilities to evaluate the change in the model, being the most common one the expected gradient length (EGL), firstly proposed in~\cite{inc:settles.nips08}, that selects the most influential sample in terms of its impact on the model parameters. Obviously, the first requirement to apply EGL is to have a learning problem tackled by gradient strategies, which is, anyhow the most usual situation. Then, the change in the model is assessed by the change in the length of the training gradient. Therefore, the instance $x_{MC}$ to be labeled is that that would result in the training gradient of the greatest value
\begin{eqnarray}
x_{\text{MC}}=\underset{x}{\text{argmax}} \sum_{i} P_{\theta} \left(y_i|x)\right) \|\nabla J_{\theta}\left(\mathcal{X},\cup \langle x,y_i\rangle\right)\|
\end{eqnarray}
where $\nabla J_{\theta} (\mathcal{X})$ is the the gradient of the objective function $J$ with respect to the parameters $\theta$ and $\| \cdot \|$ stands for the Euclidean norm. As $\|\nabla J_{\theta}(\mathcal{X})\|$ should have a value close to zero if the $J$ converged in the previous iteration, $\|\nabla J_{\theta}\left(\mathcal{X} \cup \langle x,y_i\rangle\right)\| \approx \|\nabla J_{\theta}\left(\langle x,y_i\rangle\right)\|$.
The approach based on expected error reduction puts its attention on the error committed by the model, choosing that sample that involves the greatest reduction of the error:
\begin{eqnarray}
x_{\text{ER}}=\underset{x}{\text{argmax}} \sum_{i} P_{\theta} \left(y_i|x)\right) \left(\sum_{u=1}^{\mathcal{U}} 1-P_{\theta^{+\langle x,y_i\rangle}} \left( \hat{y}|x^{(u)}\right) \right),
\end{eqnarray}
where $\theta^{+\langle x,y_i\rangle}$ stands for the new model after being trained including $\langle x,y_i\rangle$ in $\mathcal{X}$. A variation of this approach aims to minimize the variance of the model.
Both the reduction of error and the reduction of the variance analyze the whole input data set instead of individual instances. Therefore, they tend to query less outliers than strategies like USamp, QBC or EGL. An alternative way to avoid querying outliers  comes from density-based methods (DBMs), which are actually compatible and complementary to any of the approaches previously mentioned. The idea is to introduce an additional term to the query strategy that includes information about the data distribution; as a result of this, samples to be labeled are selected not only according the uncertainty or disagreement about the corresponding label but also taken into account that the sample is representative of the data set. 

\section{Numerical implementation}
Here we introduce how the AL algorithm for retrieving quantum information is implemented in order to ensure its reproducibility. We use the MATLAB Toolbox {\it Classification Learner} for a straightforward implementation, including trainers of various models, e.g., linear SVM, coarse Gaussian SVM, fine decision tree, and linear discriminator, which are selected in either USamp or QBC. We disabled validation and Principle Component Analysis before training, where other parameters are set to be the defaults. They are listed as follows for maximizing the portability:

{\bf Linear SVM} Kernal function: Linear; Kernal scale: Automatic; Box constraint level: 1 (soft margin); Standardize data: True

{\bf Coarse Gaussian SVM} Kernal function: Gaussian; Kernal scale: 5.7; Box constraint level: 1 (soft margin); Standardize data: True

{\bf Fine Decision Tree} Maximum number of splits: 100; Split Criterion: Gini's diversity index; Surrogate decision splits: Off

{\bf Linear Discriminant} Covariance structure: Full

One may notice that the hyperplane of the initial SVM is plotted either diagonally or off-diagonally in the schematic diagram of USamp, which is supposed to separate two oracles vertically. We clarify that training data are standardized by the mean of 0 and variance of 1, i.e., positions of labeled qubits are not directly used to train the model as support vectors. Thus, the support vectors for the initial SVM are $(\pm\sqrt{2}/2,\pm\sqrt{2}/2)$ or $(\pm\sqrt{2}/2,\mp\sqrt{2}/2)$, resulting in the (off-)diagonal hyperplane. However, this standardization enhances the stability of the AL algorithm against wrong labels by measurements and unpredictable perturbations of wave functions.

Codes are compatible with Matlab R2017b, in Mac OS Mojave. Codes and data are available from corresponding author upon reasonable request.

\end{document}